\documentclass[sigconf]{acmart}

\AtBeginDocument{%
 \providecommand\BibTeX{{%
   \normalfont B\kern-0.5em{\scshape i\kern-0.25em b}\kern-0.8em\TeX}}}

\usepackage{colortbl}
\usepackage{xspace}
\newcommand{\ie}{\textit{i.e., \xspace}}
\newcommand{\eg}{\textit{e.g., \xspace}}
\newcommand{\etal}{\textit{et al. \xspace}}
\usepackage{array}
\newcolumntype{L}{>{\arraybackslash}m{16cm}}
\newcolumntype{C}[1]{>{\centering\let\newline\\arraybackslash\hspace{0pt}}m{#1}}
\newcolumntype{R}[1]{>{\raggedleft\let\newline\\arraybackslash\hspace{0pt}}m{#1}}
\usepackage[numbered]{bookmark}
\usepackage{tcolorbox}
\usepackage{graphicx}
\usepackage{xcolor}
\usepackage{stfloats}
\usepackage[justification=centering]{caption}
\usepackage{dirtytalk}
\usepackage{amsmath}
\usepackage{pgfplots, pgfplotstable}
\usetikzlibrary{patterns}
\usepackage{fontawesome}

\usepackage[english]{babel}
\usepackage{textcomp}
\usepackage{mathtools}
\usepackage{ltxtable}
\usepackage{algorithm}
\usepackage{algpseudocode}
\usepackage{textcomp}

\usepackage{tcolorbox}

\usepackage{pgf-pie}
\definecolor{wedge1}{RGB}{ 190  30  46}
\definecolor{wedge2}{RGB}{ 240  65  54}
\definecolor{wedge3}{RGB}{ 241  90  43}
\definecolor{wedge4}{RGB}{ 247 148  30}
\definecolor{wedge5}{RGB}{  43  56 144}
\definecolor{wedge6}{RGB}{  28 117 188}
\definecolor{wedge7}{RGB}{  40 170 225}
\definecolor{wedge8}{RGB}{ 119 179 225}
\definecolor{wedge9}{RGB}{ 181 212 239}
\definecolor{wedge10}{RGB}{  0 104  56}
\definecolor{wedge11}{RGB}{  0 148  69}
\definecolor{wedge12}{RGB}{ 57 181  74}
\definecolor{wedge13}{RGB}{141 199  63}
\definecolor{wedge14}{RGB}{215 244  34}
\definecolor{wedge15}{RGB}{249 237  50}
\definecolor{wedge16}{RGB}{248 241 148}
\definecolor{wedge17}{RGB}{242 245 205}
\definecolor{wedge18}{RGB}{123  82  49}
\definecolor{wedge19}{RGB}{104  73 158}
\definecolor{wedge20}{RGB}{102  45 145}
\definecolor{wedge21}{RGB}{148 149 151}
\definecolor{wedge22}{RGB}{ 204 50 153}
\definecolor{wedge23}{RGB}{ 79 47 79}
\definecolor{wedge24}{RGB}{ 173 234 234}
\definecolor{wedge25}{RGB}{ 216 191 216}
\definecolor{wedge26}{RGB}{  43  56 144}
\definecolor{wedge27}{RGB}{  40 170 225}
\definecolor{wedge28}{RGB}{ 119 179 225}
\definecolor{wedge29}{RGB}{ 181 212 239}
\definecolor{wedge30}{RGB}{  0 104  56}
\definecolor{wedge31}{RGB}{  0 148  69}
\definecolor{wedge32}{RGB}{ 57 181  74}

\usetikzlibrary{chains}

\pgfmathsetmacro\startAngle{90-3.6/2}
\pgfmathsetmacro\radius{+5}
\pgfmathsetmacro\maxLeg{+12}
\pgfmathsetmacro\legBound{+60}
\pgfmathsetmacro\legSpacing{2*\legBound/(\maxLeg-1)}

\usepackage{verbatim}
\pgfplotsset{compat=1.14}
\usetikzlibrary{patterns,}

\usepackage{graphicx}
\usepackage{tabularx,booktabs}
\usepackage{array}
\usepackage{bbding}
\usepackage{pifont}
\usepackage{wasysym}
\usepackage{caption}
\usepackage{dirtytalk}
\usepackage{subcaption}
\usepackage{tabularx}
\usepackage{sfmath}
\usepackage{array, booktabs, makecell}
\usepackage{color}
\usepackage{csquotes}
\usepackage{url}
\usepackage{comment}
\usepackage{tikz}
\usepackage{balance}
\usepackage{listings}
\usepackage{booktabs} 
\usepackage{soul} 
\usetikzlibrary{fit} 
\usetikzlibrary{positioning}
\usetikzlibrary{arrows}
\usetikzlibrary{shapes.multipart}
\usepackage{adjustbox}
\usepackage{float}
\usepackage{rotating}
\usepackage{nth}
\DeclareCaptionType{TextBox}
\usepackage{pstricks}
\usepackage{placeins}
\usepackage{afterpage}
\usepackage{multirow} 
\usepackage{textcomp}
\usepackage{multicol}
\usepackage{varwidth} 

\author{Eman Abdullah AlOmar}
\affiliation{
    \institution{Stevens Institute of Technology}
    \city{Hoboken}
    \country{United States}
}
\email{ealomar@stevens.edu}

\author{Mohamed Wiem Mkaouer}
\affiliation{
    \institution{Rochester Institute of Technology}
    \city{Rochester}
    \country{United States}
}
\email{mwmvse@rit.edu}

\author{Ali Ouni}
\affiliation{
    \institution{ETS Montreal, University of Quebec}
    \city{Montreal, Quebec}
    \country{Canada}
}
\email{ali.ouni@etsmtl.ca}

\renewcommand{\shortauthors}{AlOmar et al.}

\begin{document}

\title{Automating Source Code Refactoring in the Classroom}



\begin{abstract}
Refactoring is the practice of improving software quality without altering its external behavior. Developers intuitively refactor their code for multiple purposes, such as improving program comprehension, reducing code complexity, dealing with technical debt, and removing code smells. However, no prior studies have exposed the students to an experience of the process of antipatterns \textit{detection} and refactoring \textit{correction}, and provided students with toolset to practice it. To understand and increase the awareness of refactoring concepts, in this paper, we aim to 
 reflect on our experience with teaching refactoring and how it helps students become more aware of bad programming practices and the importance of correcting them via refactoring. This paper discusses the results of an experiment in the classroom that involved carrying out various refactoring activities for the purpose of removing antipatterns using JDeodorant, an Eclipse plugin that supports antipatterns detection and refactoring.  
 The results of the quantitative and qualitative analysis with 171 students 
  show that students tend to appreciate the idea of learning refactoring and are satisfied with various aspects of the JDeodorant plugin's operation. Through this experiment, refactoring can turn into a vital part of the computing educational plan. We envision our findings enabling educators to support students with refactoring tools tuned towards safer and trustworthy refactoring.

\end{abstract}





\maketitle
\section{Introduction}
\label{Section:Introduction}

Design antipatterns are symptoms of poor choices at the software architecture level. These bad programming practices typically violate object-oriented design principles, such as \textit{Single Responsibility} and \textit{Law of Demeter}. 
  The existence of these design antipatterns often leads to the degradation of software architectures, making them difficult to understand, reuse, and evolve. It is important to note that these antipatterns are different from coding errors, and do not directly lead to compiler or logical faults, but various studies have demonstrated how the existence of antipatterns makes the code significantly more prone to errors \cite{bessghaier2021longitudinal,khomh2009exploratory,khomh2012exploratory}.

Two popular examples of design antipatterns are \texttt{God Class} and \texttt{Feature Envy}. The first characterizes classes that are abnormally large and monopolize most of the system's behavior by controlling a significant number of other coupled classes. Decomposing this class is trivial to sustain the modular design of the system. The second is related to methods that heavily rely on methods and attributes that are outside of its class more than those inside it. This is a symptom of a misplaced method that needs to be moved to a class to make it more cohesive.
 
 
 To cope with these antipatterns, refactoring has emerged as a \textit{de-facto} practice to improve software quality through the removal of antipatterns \cite{cunningham1992wycash}. Refactoring is the art of improving source code internal design, without altering its external behavior \cite{Fowler:1999:RID:311424,alomar2021preserving}. 


Several studies have proposed methods for teaching code refactoring through the identification of duplicate and dead code, and bad naming conventions \cite{keuning2021tutoring,haendler2019interactive,keuning2020student,haendler2019refactutor}. While these techniques play a role in improving the student's understanding of refactoring, it is critical to expose students to deeper design-level antipatterns that are frequently found even in well-engineered projects \cite{palomba2018diffuseness}, and harder to fix \cite{bavota2015experimental}. For instance, early exposure to God Classes and Feature Envies would help reduce their prevalence in the future.

Therefore, the \textit{goal} of this paper is to increase awareness of bad programming practices, \ie design-level antipatterns, and the importance of correcting them through the application of appropriate refactoring operations. Hence, we perform a series of assignments where students are asked to reason over how to refactor the \texttt{God Class} and \texttt{Feature Envy} design antipatterns. We chose these specific antipatterns on the basis of their frequent refactoring by developers in various systems\footnote{Based on tool usage statistics: https://users.encs.concordia.ca/~nikolaos/."}. 

We report our experience using JDeodorant \cite{tsantalis2018ten}, an integrated development environment (IDE) plugin, to support students in finding suitable refactorings. We chose JDeodorant because it is widely used by researchers as the state-of-the-art benchmark to assess the precision of refactoring techniques. JDeodorant is also widely adopted by practitioners to improve their systems' design.

We adopted the reflective learning strategy when designing the refactoring assignments \cite{brockbank2007facilitating}. In fact, we follow Ash and Clayton's \textit{DEAL} model \cite{ash2009generating} as we aim to let students first construct and \textit{describe} an initial refactoring solution before \textit{examining} other candidate solutions recommended by JDeodorant, to finally \textit{articulate} on the difference between their solution and the ones recommended by the tool. We executed these assignments in undergraduate and graduate software engineering courses at two universities\footnote{Hidden for double-blind review.}. We analyzed 171 student refactoring submissions in terms of two dimensions. The first dimension is empirical, as we assess the quality of students' refactored code in contrast with JDeodorant's, to extract any knowledge gap. This dimension's outcome reveals how \texttt{God class} antipattern tends to be harder for students to refactor, compared to \texttt{Feature Envy}, and how the use of JDeodorant has facilitated the correction of these hard instances. The second dimension is qualitative, where we survey students to sense their feedback on the tool's usefulness, usability, and functionality. The results of the survey show that the vast majority of students (87\% responses) found the plugin to be useful, and usable, and were satisfied with its operation. Finally, we reflect on the importance of reinforcing software design principles and patterns. Therefore, we foresee students' usage of JDeodorant as an opportunity to raise awareness of the detection of antipatterns and their correction measures.

This paper contributes to the broad adoption of refactoring by (i) designing practical assignments that first challenge students' abilities to refactor design-level problems, then second provide them with candidate solutions to reason over and choose based on their potential impact on quality, and (ii) reporting the experience of using the JDeodorant plugin. This experiment enabled instructors to design personalized, hands-on assignments and support students in learning how to use refactoring features in the IDE. It also achieves another learning objective, since a recent study has shown that developers rarely use the built-in IDE features when refactoring their code, increasing the error proneness of their changes \cite{golubev2021one}. As part of this paper's contributions, we provide the assignment description, the tool documentation, and statistical tests for educators to replicate and extend \cite{JDeodorant-replication}.


\vspace{-.3cm}
\section{JDeodorant Workflow}
\label{Section:workflow}

JDeodorant \cite{tsantalis2018ten} stands as one of the popular refactoring tools that have been provided as an Eclipse and IntelliJ IDEA plugins. It automatically detects antipatterns, including \texttt{Feature Envy}, \texttt{State Checking}, \texttt{Type Checking}, \texttt{Long Method}, \texttt{God Class}, \texttt{Duplicate Code}, and \texttt{Refused Bequest}, and for each detected antipattern, it offers its correction by providing a list of candidate refactorings that developers have to choose from. Developers, are then responsible for choosing the most adequate refactoring operations according to their design choices and preferences. 

\begin{figure}[t]
\centering 
\includegraphics[width=\columnwidth]{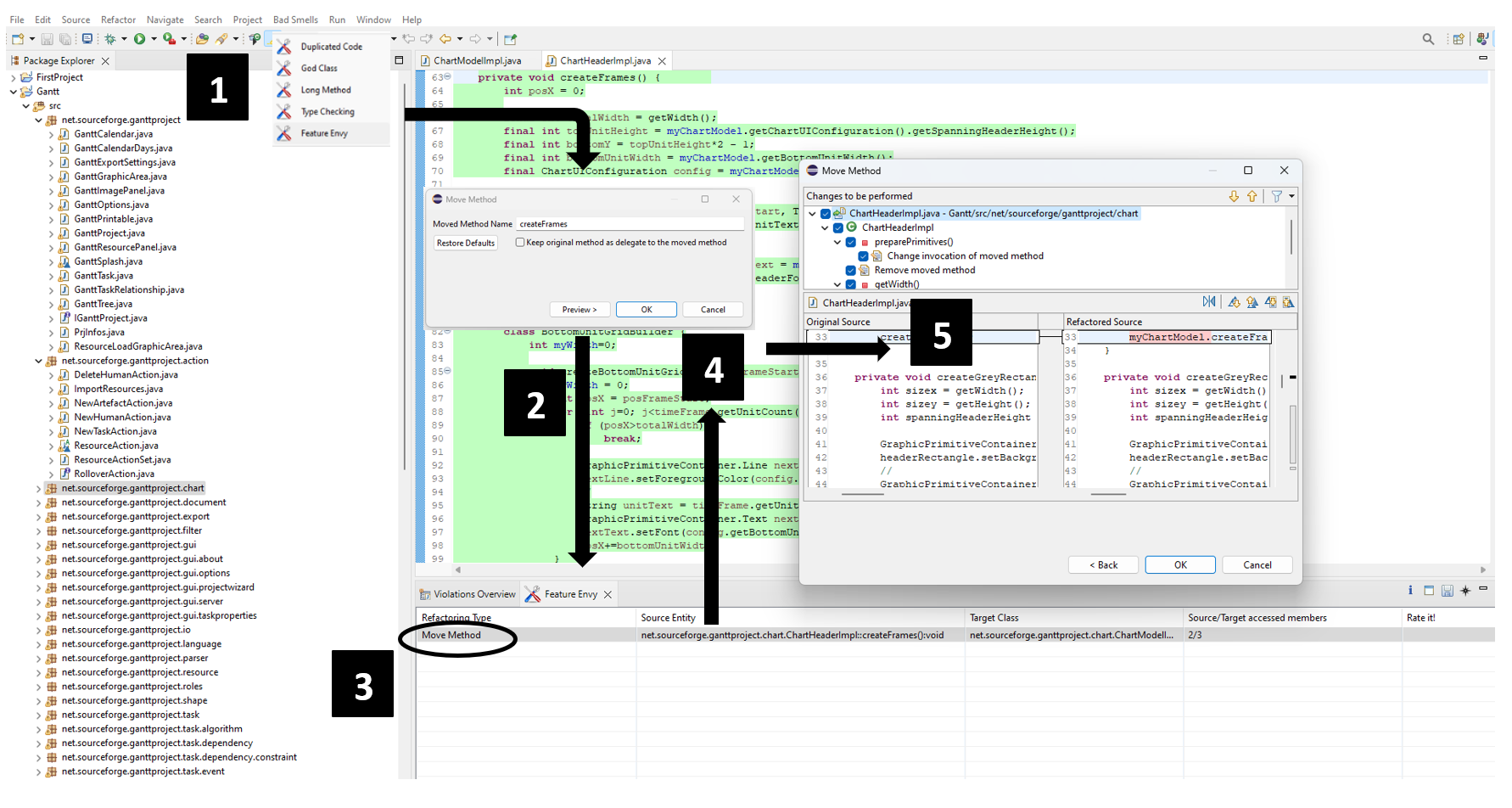}
\caption{JDeodorant workflow.}
\label{fig:jdeodorant-workflow}
\end{figure}

To illustrate the workflow of JDeodorant, we choose to fix the \texttt{Feature Envy} antipattern that may exist in the Gantt\footnote{https://github.com/bardsoftware/ganttproject} project. We open the Gantt project using the Eclipse IDE, with JDeodorant already installed as a plugin. Then, we enable the plugin by clicking on \texttt{Bad Smells} in the menu bar (Step 1), and then we click on \texttt{Feature Envy}. To identify all instances of this antipattern, the plugin internally parses all the project methods to generate their corresponding Abstract Syntax Tree (AST) representations, which in turn allows the tool to determine whether a method matches the pattern of a \texttt{Feature Envy} antipattern. Once the detection process is done, all \texttt{Feature Envy} instances are shown in the plugin view (Step 2), along with their corresponding refactoring suggestions (Step 3). For example, as can be seen in Figure \ref{fig:jdeodorant-workflow}, the \texttt{createFrame} method, with 55 lines of code, and located in the class \texttt{CharHeaderImpl}, is flagged as a \texttt{Feature Envy}. The plugin also proposed a couple of candidate \texttt{Move Method} Refactorings for us to choose from. If we choose to fix the \texttt{creatFrame} method by selecting the first refactoring suggestion, the plugin displays the method being refactored (Step 4), and internally calls the \texttt{Move Method} built-in feature inside Eclipse. This feature initiates the refactoring process by opening a preview window (Step 5). This window shows the original code of the \texttt{createFrame} method, side by side with a preview of refactored code, which would show the result of moving \texttt{createFrame} to another class. If we confirm the refactoring, it will automatically be applied to the source code.

\section{Experimental Setup}
\label{Section:Assignment}



\subsection{Teaching Context and Participants}

The study is performed in courses taught at 2 universities\footnote{Hidden for double-blind review. Details will be added upon the paper's acceptance.}. The courses cover various concepts related to software analysis and testing, along with practical tools, widely used in the open-source community. Students were also given several hands-on assignments in topics including software quality metrics, code refactoring, bug management, unit and mutation testing, and technical debt. Before conducting the experiment, students acquired the necessary background by learning the following concepts: (1) code quality (teaching quality and how to measure it), (2) design antipatterns (teaching violations of design principles, and their detections rules), and (3) code refactoring (teaching refactoring recipes and operations). The experiment's assignments constituted 7.5\% of the final grade. It was due 14 days each after the concepts were taught. 

\subsection{Assignments Content and Format}

We adopt Ash and Clayton's reflection model by gathering evidence (refactored code) that can be examined to identify any gaps in the state of refactoring practice (inability to correct certain antipatterns), with the intent to improve it (provide alternative correction mediums).


Initially, students are asked to analyze one version of a Java software system of their choice approved by the instructor to ensure its eligibility based on popularity, besides making sure it correctly compiles since JDeodorant requires it. The rationale behind giving students the choice of project, is to let them choose one that they are comfortable with, and fits into their interests. 
 For students who do not want to search for a project, they are given a list that the instructor has already curated. 
We selected these projects \cite{JDeodorant-replication} because they contain the antipatterns that we are interested in. 
We conduct our experiments through two assignments: In the first assignment, students are asked to fix the two antipatterns (\ie \texttt{God Class} and \textit{ii}) \texttt{Feature Envy}) and provide a sequence of refactoring operations that will fix multiple instances of these antipatterns. The submissions to this assignment constitute the \textit{Manual Refactoring}. In the second assignment, students are asked to set up and run JDeodorant to analyze the chosen project production code. Upon running JDeodorant, students are required to choose at least 2 antipatterns instances from 2 different antipatterns types supported by the tool (4 in total), and then analyze JDeodorant recommendations to choose the potential refactoring operations to fix them. Since JDeodorant gives many recommendations on how to fix the same antipatterns instance, based on the students' understanding of problems' symptoms, they would need to reason when choosing the code changes that remove the smells while fitting properly in the system's design. The submissions to this assignment constitute the \textit{Assisted Refactoring}. 
 In summary, the students followed these steps:

\begin{enumerate}
    \item Manually fix the detected antipatterns types: (\textit{i}) \texttt{God Class}, and (\textit{ii}) \texttt{Feature Envy}.
    \item Provide a sequence of refactoring operations that will fix multiple instances of those antipatterns (\ie \textit{Manual Refactoring}).
    \item Justify the choices regarding refactoring decisions for each fixed antipattern type. 
    \item Install the Eclipse plug-in for JDeodorant.
    \item Run JDeodorant on a project of students' choice and select 2 instances of each of the 2 following antipatterns types: (\textit{i}) \texttt{God Class}, and (\textit{ii}) \texttt{Feature Envy}.
    \item Look at the refactoring recommendations by JDeodorant, and choose which ones to be executed. Students keep refactoring until processing all their chosen smell instances.
    \item Report the findings: chosen antipattern instances, chosen refactoring operations and results (\ie \textit{Assisted Refactoring}).
    \item Add to the report a concise comment about the experience with JDeodorant (Optional).
    
\end{enumerate}


Students were evaluated based on two aspects, (1) concept understanding: assessment of students’ ability to apply the right refactoring to fix antipatterns; (2) program analysis: assessment of whether students are able to execute refactoring and verify the preserved behavior. Students were not evaluated on their perception of the code, to avoid any cognitive bias that may occur under the pressure of being graded. Also, we anonymized the feedback, and made it optional, to only collect information from students who were serious about it, which will increase the magnitude of provided experience. Despite it being not mandatory, the majority of students (96.07\%) chose to complete it. 

The assignment was performed over four consecutive semesters. 
 171 students, primarily from computer science (CS) and software engineering (SE) majors, were enrolled during these semesters, and completed assignments.

\subsection{Data Analysis}

We analyzed the responses to open-ended questions to create a comprehensive high-level list of themes by adopting a thematic analysis approach based on guidelines provided by Cruzes~\etal~\cite{cruzes2011recommended}. Thematic analysis is one of the most used methods in Software Engineering literature~\cite{Silva:2016:WWR:2950290.2950305,alomar2022code,alomar2023use}. This is a technique for identifying and recording patterns (or \say{themes}) within a collection of descriptive labels, which we call \say{codes}. For each response, we proceeded with the analysis using the following steps: \textit{i}) Initial reading of the survey responses; \textit{ii}) Generating initial codes (\ie labels) for each response; \textit{iii}) Translating codes into themes, sub-themes, and higher-order themes; \textit{iv}) Reviewing the themes to find opportunities for merging; \textit{v}) Defining and naming the final themes, and creating a model of higher-order themes and their underlying evidence. The above-mentioned steps were performed independently by
two authors. One author performed the labeling of students' comments independently from the other author who was responsible for reviewing the currently drafted themes. By the end of each iteration, the authors met and refined the themes. 



\section{Results}
\label{Section:Result}

\begin{figure*}[t]   
\centering
\begin{subfigure}{0.4\linewidth}
\centering
\includegraphics[width=0.5\linewidth]{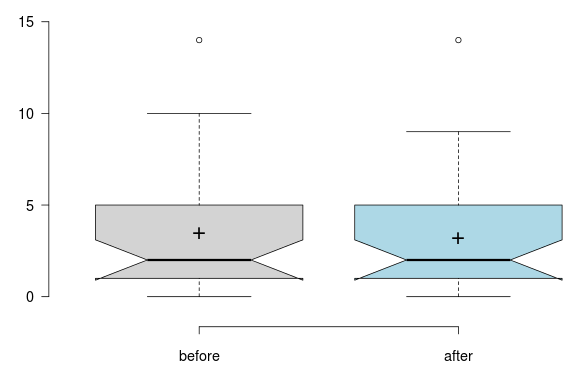}    
\caption{Impact of Manual Refactoring on God Class Distribution}
\label{BP:God Class}
\end{subfigure}%
\begin{subfigure}{0.4\linewidth}
\centering
\includegraphics[width=0.5\linewidth]{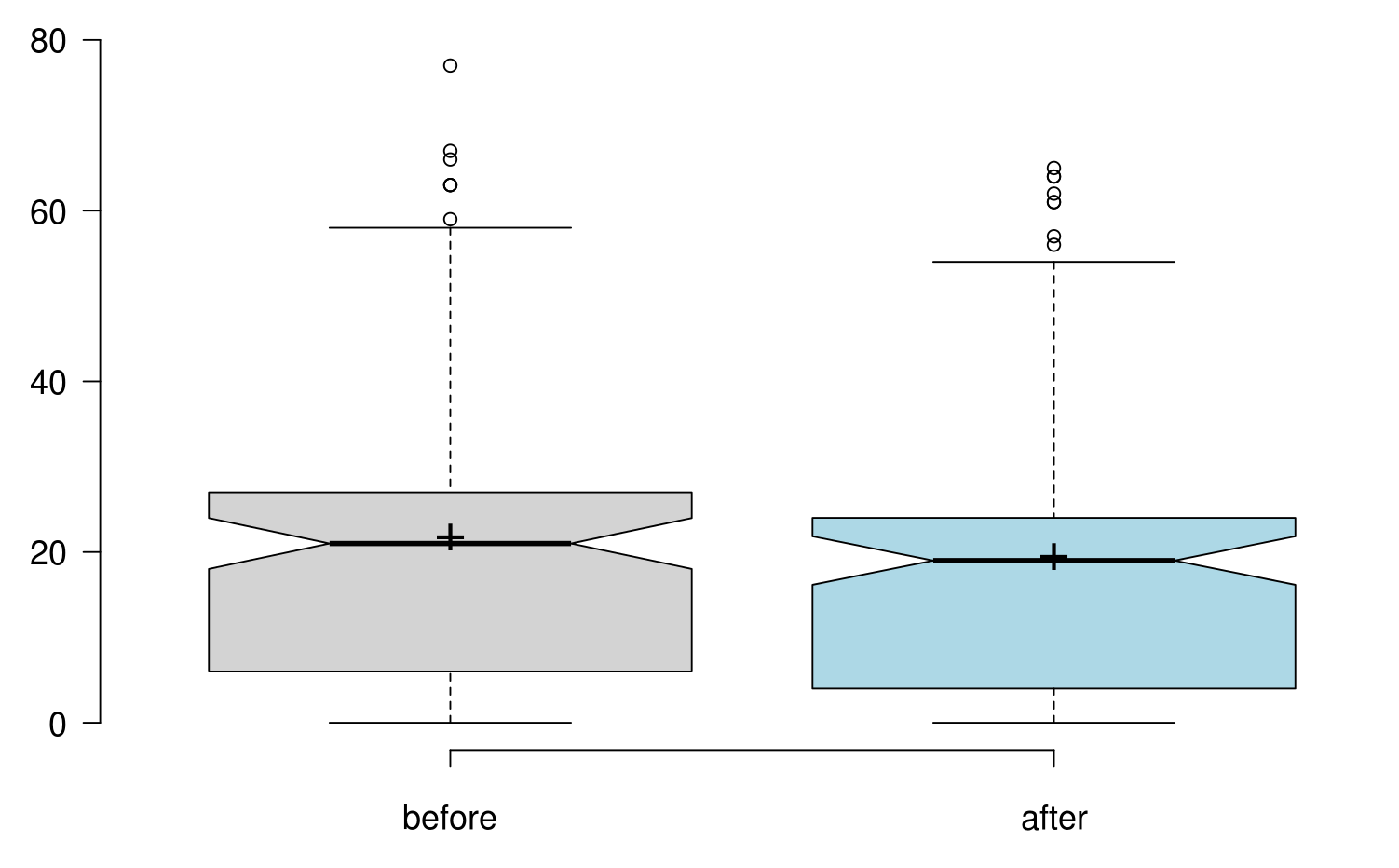}   
\caption{Impact of Assisted Refactoring on God Class Distribution}
\label{BP:God Class in infusion}
\end{subfigure}
\centering
\begin{subfigure}{0.4\linewidth}
\centering
\includegraphics[width=0.5\linewidth]{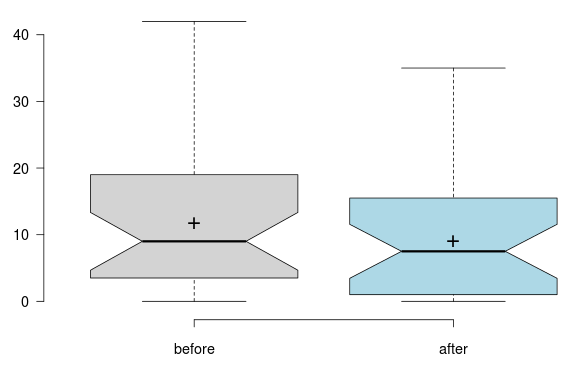}    
\caption{Impact of Manual Refactoring on Feature Envy Distribution}
\label{BP:God Class-manual}
\end{subfigure}
\begin{subfigure}{0.4\linewidth}
\centering
\includegraphics[width=0.5\linewidth]{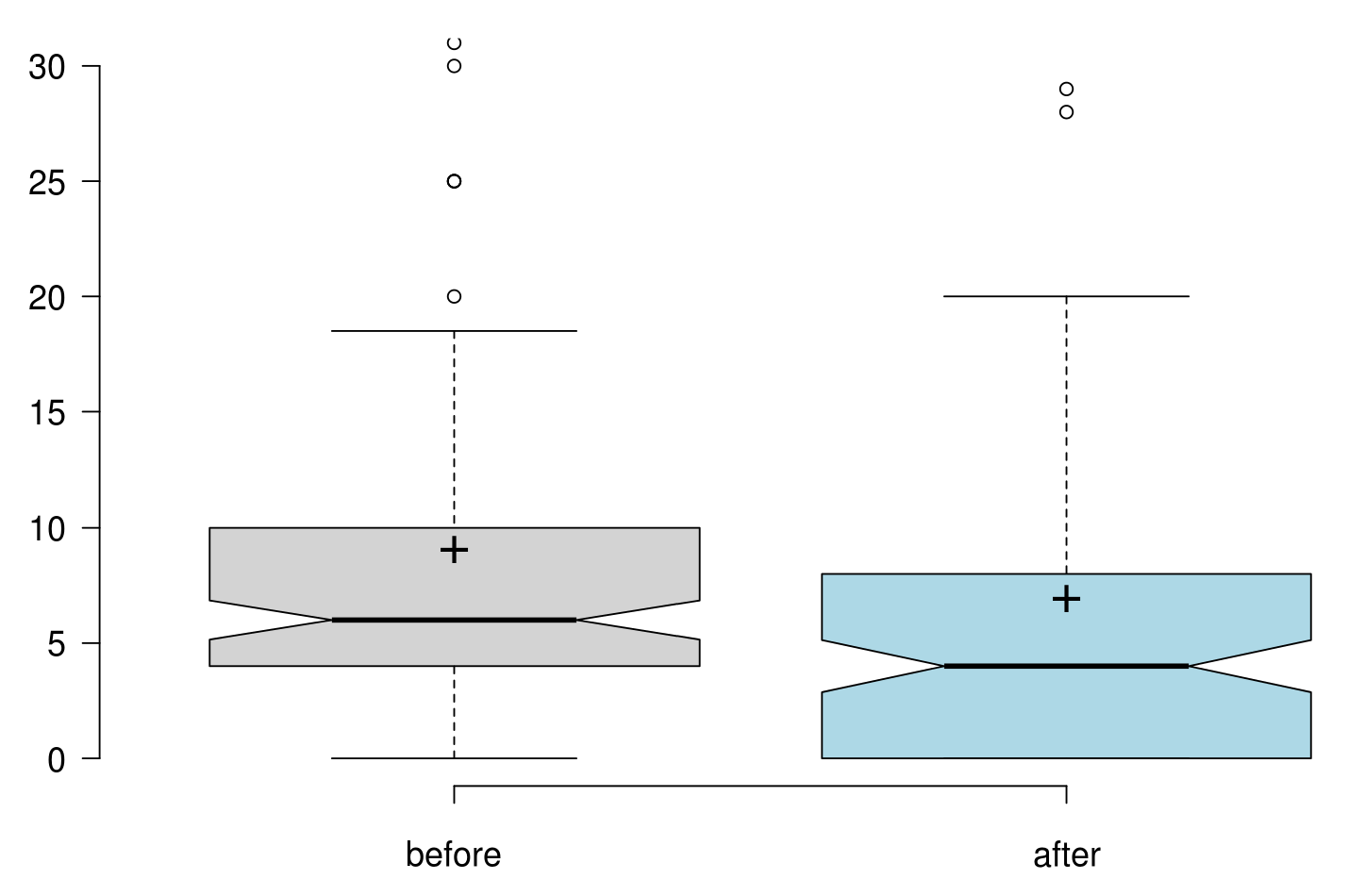}   
\caption{Impact of Assisted Refactoring on Feature Envy Distribution}
\label{BP:Feature Envy-manual}
\end{subfigure}
\caption{\textcolor{black}{Boxplots of (a) God Class, and (b) Feature Envy antipattern instances addressed by students.}} 
\label{Chart:Boxplots}
\vspace{-.3cm}
\end{figure*}

\begin{figure}[t]
 	\centering
 	\includegraphics[width=1.0\columnwidth]{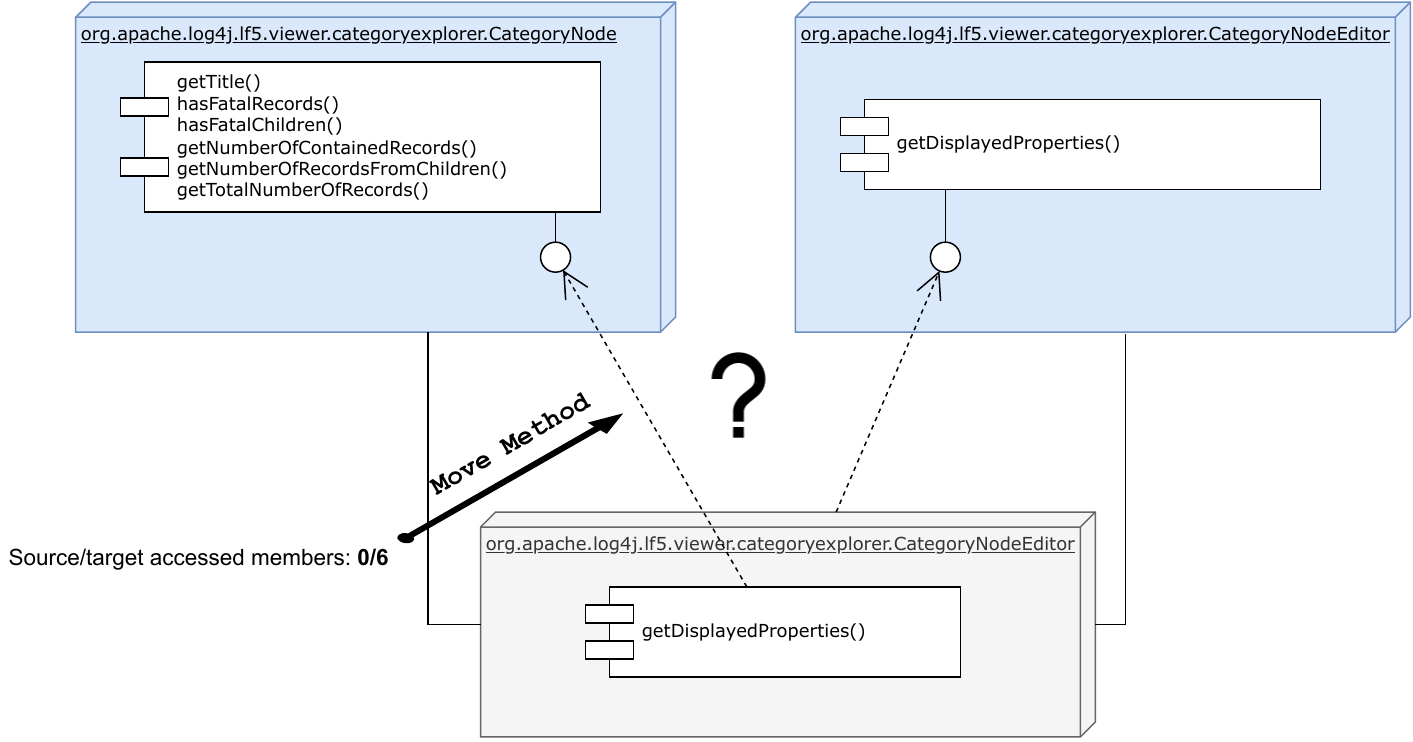}
 	\caption{\texttt{Feature Envy} example from the Log4J project \cite{Log4J}.}
 	\label{fig:example}
\end{figure}

\subsection{Quantitative Analysis}
\begin{table}[H]
  \centering
\caption{Statistical test.}
\label{Table:statistical test}
\begin{adjustbox}{width=0.48\textwidth,center}
\begin{tabular}{lllll}\hline
\toprule
\bfseries Antipattern & \bfseries Approach & \bfseries Impact & \bfseries \textit{p}-value & \bfseries Cliff's delta ($\delta$) \\
\midrule
\texttt{God Class} &  Manual  & +ve & \textit{\textbf{0.01}} & 0.05 (Negligible) \\
& Assisted & +ve & \textit{\textbf{1.89e-161}} & 0.76 (Large)
\\ \hline
\texttt{Feature Envy} &  Manual  & +ve & \textit{\textbf{2.22e-06}} & 0.19 (Small)
\\
         &  Assisted & +ve & \textit{\textbf{1.46e-161}}  & 0.76 (Large)
\\
\bottomrule
\end{tabular}
\end{adjustbox}
\end{table}
\textcolor{black}{To show the effectiveness of JDeodorant in educating students about better-making design decisions, we count the number of \texttt{God Class}, and \texttt{Feature Envy} antipatterns before and after refactoring, initially when students refactored the antipatterns manually (referred to as \textit{Manual Refactoring}), and then when students refactored the antipatterns based on JDeodorant recommendations (referred to as \textit{Assisted Refactoring}). Figure \ref{Chart:Boxplots} reports the box plots depicting the distribution of each group value, clustered by the two above-mentioned antipatterns. We used the Wilcoxon test \cite{wilcoxon1945individual} to test the significance of the difference between the group's values. This non-parametric test checks continuous or ordinal data for a significant difference between two dependent groups. Our hypothesis is formulated to test whether the antipattern after the refactoring group is significantly lower than the values of the antipatterns before the refactoring group. The difference is considered statistically significant if the \textit{p}-value is less than 0.05. Furthermore, we used Cliff's delta ($\delta$) effect size to estimate the magnitude of the differences. Regarding its interpretation, we follow the guidelines reported by \cite{trove.nla.gov.au/work/16432558}:  Negligible for $\mid \delta \mid< 0.147$; Small for $0.147 \leq \mid \delta \mid < 0.33$; Medium for $0.33 \leq \mid \delta \mid < 0.474$; and  Large for $\mid \delta \mid \geq 0.474$.}

 \textcolor{black}{The count of corrected antipatterns was done manually by the authors. Since the complete list of students' submissions is too large to examine manually, we selected a statistically significant sample for our analysis and annotated 62 submissions. This quantity roughly equates to a sample size with a confidence level of 95\%, with a margin of error of 10\%. 
As can be seen in Figure \ref{Chart:Boxplots}, antipatterns (\ie \texttt{God Class} and \texttt{Feature Envy}) before refactoring are larger than the antipatterns after refactoring. While the difference was statistically significant (0.011 and 2.22e-0 for \texttt{God Class} and \texttt{Feature Envy}, respectively), the magnitude of the difference is negligible and small for \texttt{God Class} and \texttt{Feature Envy}, respectively. We conjecture that although there is quality improvement as the number of antipatterns decreased, there were many instances where students' \textit{Manual Refactorings} could not remove certain antipatterns, particularly the God Classes instances.  
 However, \textit{ assisted refactoring} also had a positive impact on quality, as the number of antipatterns (\ie \texttt{God Class} and \texttt{Feature Envy}) before refactoring is greater than the antipatterns after refactoring (1.89e-161 and 1.46e-161 for \texttt{God Class} and \texttt{ feature envy}, respectively). However, the main difference lies in the magnitude of the difference (Cliff Delta), which was large for the \textit{Assisted Refactoring}, according to Table \ref{Table:statistical test}. We conclude that JDeodorant's assistance was beneficial in improving students' design decisions. 
 Figure \ref{fig:example} shows an example of a \texttt{Feature Envy} design antipattern from one of the student's submissions. The \texttt{CategoryNodeEditor} class has a method called \texttt{getDisplayedProperties} that seems overly interested in the properties of the \texttt{CategoryNode} class. The method \texttt{getDisplayedProperties} calls many methods from \texttt{CategoryNode} class more than its own class methods (Coupling = 6). This indicates that the method should belong to the \texttt{CategoryNode} class instead. When the student applied the \texttt{Move Method} refactoring, as recommended by JDeodorant, the \texttt{Feature Envy} antipattern was removed, along with a decrease in the system's overall coupling.} 

\noindent\textbf{Observation. \textit{Not all antipatterns are easily refactorable.}} It is important to note that despite students' efforts to remove antipatterns, our analysis of \textit{Manual Refactoring} code shows how many instances of antipatterns existed after the refactoring session. In particular, God Classes are difficult to refactor, as the magnitude of the difference is negligible in Table \ref{Table:statistical test}. Previous research has demonstrated how God classes tend to be hard to fix in industry \cite{anquetil2019decomposing}. So, we emphasize the importance of raising students' awareness, as a preventive measure, to avoid creating God classes.
\begin{table*}[ht]
  \centering
	 \caption{\textcolor{black}{Student's insight about the usefulness,  usability and functionality of the tool.}}
	 \label{Table:example}
\begin{adjustbox}{width=1.0\textwidth,center}
\begin{tabular}{llLllll}\hline
\toprule
\bfseries Theme & \bfseries Sub-theme  & \bfseries Example (Excerpts from a related student's comment) \\
\midrule
\multirow{8}{*} {\textbf{Usefulness}} & \cellcolor{gray!30}{Efficiency}  & \cellcolor{gray!30} \say{\textit{JDeodorant is undoubtedly a very convenient tool for developers, he can easily fix various code smells and \textbf{improve the development efficiency} of developers.}} \\ 
&  Quality & \say{\textit{It can help to prioritize refactoring instances by selecting them based on their number of methods 
and attributes they have. It also gives more details about classes and have an \textbf{insight of the impact 
of refactoring}.}} \\ 
& \cellcolor{gray!30}{Automation} & \cellcolor{gray!30} \say{\textit{Overall my experience with the plugin was good.  It takes most of the refactoring work and \textbf{basically automates it}.}} \\ 
& Awareness &  \say{\textit{The product seemed to work well. Additionally, its ability to tell what refactoring to do and actually perform it seems super helpful, and \textbf{more helpful if I actually knew what I was doing when refactoring}.}} \\ 
& \cellcolor{gray!30}{Experience} &\cellcolor{gray!30}  \say{\textit{My experience with JDeodorant was quite pleasant. It \textbf{simplified the process} of addressing flaws 
in the code and \textbf{for someone with limited software engineering experience} related to code smells it was very informative.}} \\
\hline
\multirow{7}{*} {\textbf{Usability}} &  Graphical design &  \say{\textit{the very intuitive \textbf{graphical design} of the plugin (for instance, doing 
things such as highlighting code that it changes in green) \textbf{makes such reviews far simpler} than alternative approaches may provide.}} \\ 
&  \cellcolor{gray!30}Preview & \cellcolor{gray!30} \say{\textit{I particularly liked
IntelliJ’s display \textbf{showing side-by-side diffs of what changed}; it made it clear what was changing, which in turn
made it easier to interpret why the tool was suggesting the change.}} \\ 
&  Visualization &  \say{\textit{The JDeodorant version for eclipse \textbf{provides the visual
context/flow chart} containing the breakdown of the bad smells and the highlighted
code. That proved to be very beneficial during the refactoring process, especially with
god class.}} \\ 
& \cellcolor{gray!30}Documentation & \cellcolor{gray!30} \say{\textit{It is also an automated process, which is User
friendly and \textbf{it gives guidelines for the usage} and it also pre-evaluates the refactoring.}} \\
\hline
\multirow{3}{*} {\textbf{Functionality}} & Design antipatterns &  \say{\textit{I found that jDeodorant did a \textbf{great job at detecting any issues} that were found.}} \\ 
& \cellcolor{gray!30}Refactoring & \cellcolor{gray!30} \say{\textit{I am happy I picked a smaller project to work on, it is 
significantly easier to see the effects of the changes. Most of the \textbf{refactoring that was suggested were 
various kinds of extraction}, mostly methods.}} \\
\hline
\multirow{8}{*} {\textbf{Recommendation}} &  Quality &  \say{\textit{It
would have been nicer if the project \textbf{checked for refactorings} again automatically \textbf{after applying}
one instead of the user needing to use it.}}  \\
& \cellcolor{gray!30}Design antipatterns & \cellcolor{gray!30} \say{\textit{I am really impressed with how this tool can detect the code 
smells and suggest refactoring to solve the problems. However, [...], it 
can \textbf{only detect three types of code smells}. Many software has a wide range of other code smells not only 
(long method, god class, feature envy) which are not possible to detect using this tool. }}\\
& Refactoring & \say{\textit{some of it's \textbf{suggestion of the refactoring is
incompleted}, the suggestion name and logic detection has missed some part of the code. So I feel
this tool can only use as an quick reference in the code quality review}} \\
& \cellcolor{gray!30}Testing & \cellcolor{gray!30} \say{\textit{Its \textbf{hard to find} if refactoring did \textbf{break any of the feature} or functionality remains
the same even after refactoring.}} \\

\bottomrule
\end{tabular}
\end{adjustbox}
\vspace{-.3cm}
\end{table*}

\noindent\textbf{Observation. \textit{Understanding the impact of refactoring on quality is challenging for students.}} Although both refactoring sessions aim to improve quality, students realize, when they compare their manual refactoring with the assisted one, that not all refactorings can be equally beneficial to design quality. For instance, the process of extracting multiple classes, to remove a \textit{God Class} antipattern, will eventually increase the number of classes per package, which is also considered an increase in system complexity according to the CK quality metrics \cite{chidamber1994metrics}. Thus, students need to consider these trade-offs as they make their choices. In this context, quality gate tools, such as \textsc{Understand}\footnote{\url{https://scitools.com/}} and SonarQube\footnote{https://sonarqube.org} can be deployed to measure the quality before and after the application of refactoring. This might strengthen students' understanding of writing well-structured code and raise their confidence to perform the recommended refactoring.

 


\vspace{-.4cm}
\subsection{Qualitative Analysis}

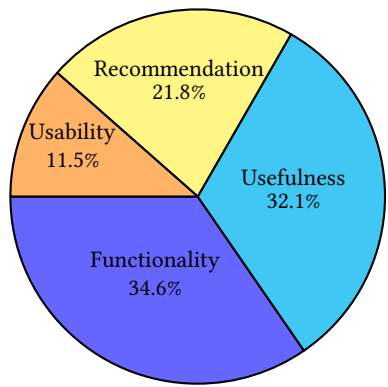
\begin{figure}
\centering 
\begin{tikzpicture}
\begin{scope}[scale=0.83]
\pie[rotate = 180,pos ={0,0},text=inside,no number]{34.6/Functionality\and34.6\%, 32.1/Usefulness\and32.1\%, 21.8/Recommendation\and21.8\%,11.5/Usability\and11.5\%}
\end{scope}
\end{tikzpicture}
\caption{Students' perception about the refactoring tool.} 
\label{fig:motivation}
\vspace{-.6cm}
\end{figure}

In Table~\ref{Table:example}, we report the main thoughts, comments, and suggestions about the overall impression of the usefulness, usability, functionality, and recommendation of the tool, according to the conducted labeling. Figure \ref{fig:motivation} shows the percentages of students' insight. As can be seen, the `Functionality' and `Usefulness' categories had the highest number of responses, with a response ratio of  34.6\% and 32.1\%, respectively. The category `Recommendation' was the third most popular category with 21.8\%, followed by `Usability', which had a ratio of 11.5\%. This finding indicates what students mainly care about when using the tool. Table~\ref{Table:example} presents samples of the students' comments to illustrate their impression of each theme.

\textbf{Usefulness.} Generally, the respondents found the plugin to be useful in regards to five main aspects: efficiency, quality, automation, awareness, and experience. 40 out of 171 
students commented that JDeodorant is very intuitive to use and was quite efficient to find refactoring opportunities, and convenient for developers who would not have to examine and refactor antipatterns manually. 30 students communicated that eliminating antipatterns assists in increasing its \textit{readability} while reducing its \textit{coupling} and \textit{complexity}, which helps improve overall code quality. A few students revealed that the tool's ability to identify antipatterns within a selected file allows them to only correct errors in a specific location/file of their interest, instead of inspecting the entire project. Further,  two students commented that the tool aids less experienced developers in identifying the antipatterns when updating a source file that they are not necessarily familiar with. Additionally, a group of students mentioned that the tool helps less experienced and novice developers in writing well-structured code. 
 
\textbf{Usability.}  Based on the feedback provided by the students, the key areas in usability related to the graphical design, preview, visualization, and documentation. Five students pointed out the graphical design of the plugin is intuitive, especially the IntelliJ IDEA's display feature showing side-by-side diffs of what changed. This preview feature makes it clear what was changing and why the tool was suggesting the change. Other comments also stated the importance of the preview feature, which allows them to foresee the impact of the change without actually performing it. Two students reported the useful feature of antipattern visualization as a flow chart, as it allows locating the \textit{hot} areas in code that encapsulate a large set of smells. Lastly, the documentation of the plugin is written and easy to follow. 

\textbf{Functionality.} According to the students' feedback about the tool's functional features, 34.6\% of the students' comments show a couple of appreciation for the supported antipatterns by the tool, and how this feature helps in better understanding the concepts in a real-world scenario. Additionally, the students commented on their ability to practice a variety type of refactoring operations according to their removal of the antipatterns. 

\textbf{Recommendation.} From the students' feedback, we have also extracted suggestions to improve the tool's features. 21.8\% of the students' comments show a couple of suggested changes as a recommendation or refactoring support to be made to the tool's operation or UI. We found out the students pointed out some of the recommendations related to quality, antipatterns, refactoring, and testing. Students recommend the tool to perform a sanity check after performing the application of refactoring, support more antipatterns, perform the complete application of refactoring, and perform testing in order to check the behavior preservation of code transformation after refactoring.


\section{Reflections}
\label{Section:Reflection}



\faThumbTack\noindent\textbf{ Reflection \#1:  \textit{Raising the awareness of antipatterns and their refactoring strategies.}}
Previous studies \cite{smith2006innovative,stoecklin2007teaching,abid2015reflections,izu2022resource,lopez2014design} have proposed different methods to teach refactoring, but did not provide students with toolset to practice it. 
 It is still a manual process that might contribute in a limited way in the long run. However, if students learn how to automate refactoring, it helps them to better refactor the code while preserving the behavior and making the code less prone to error. Further, in education, there are many concepts related to design principles that are taught in the classroom, such as SOLID and GRASP. 
 However, the responses that we received from students, have shown the importance of covering design anti-patterns, and bad programming practices avoidance in curricula, as these topics are generally less popular, compared to design patterns. 
 In many CS and SE curricula, instructors highlight software quality when teaching good programming practices and design patterns. This assignment complements it by revealing how deviation from best practices can lead to poor design choices that negatively impact the source code.  
  Similarly to teaching blueprints of design patterns, students should also be exposed early to quality concerns and encouraged to improve the design of their own code.
Moreover, one of the noteworthy points is that using the tool in the assignment helps less experienced and novice coders to write well-structured code. To further raise awareness, educators can include empirical evidence to enhance students' understanding of refactoring and antipatterns concepts and so students providing low-quality code can be convinced that these concepts help improve the quality of software systems.

\faThumbTack\noindent\textbf{ Reflection \#2: \textit{Reinforcing software engineering principles and good development practices.}}
Studies have shown how group-based project artifacts tend to be purposely over-engineered complex models as a means to showcase the ability to design complex systems \cite{colbeck2000grouping,berland2013student}. Through this assignment, students would realize that over-engineered systems tend to contain more design anti-patterns, and therefore simplifying them is necessary for their code to become easier to maintain in the future.
According to students' insights, we believe that instructors need to highlight the desirable properties of refactoring tools (\eg quality improvement, developer perception, automated testing, etc). Future educators and researchers are encouraged to revisit existing refactoring tools so that students can gain more confidence in using them. Moreover, since the classical definition of refactoring focuses on behavior preservation of the applied transformation, instructors might consider pointing out some behavior preservation strategies (\eg \cite{alomar2021preserving}) and explore their potential in assessing the correctness of the refactored code (\eg for the context of JDeodorant, the students can run unit tests to verify that the applied refactoring was behaviorally preserved).

\vspace{-.4cm}
\section{Related Work}
\label{Section:Background}

 \textcolor{black}{Smith \etal and Stoecklin \etal \cite{smith2006innovative,stoecklin2007teaching}  introduced an incremental approach by focusing on lessons from an
innovative pedagogical approach to teaching refactoring, such as self-documenting code and better recognizing code. The authors conclude that refactoring can become an integral component of the computer science curriculum by reinforcing software engineering principles. Rabb \cite{raab2012codesmellexplorer} introduced  CodeSmellExplorer to familiarize users
with good coding practices by visualizing an interactive graph network of antipatterns and connected refactorings. Lobez \etal \cite{lopez2014design} described e-activities for teaching refactoring by following Bloom’s
taxonomy (\ie proposing activities to help with understanding a concept, applying refactorings in the
context of and synthesizing of the use of refactorings in open source projects). 
Elezi \etal \cite{elezi2016game} proposed a gamification system
that tracks and rewards refactorings during development. Haendler and Neumann \cite{haendler2019serious} explored the challenges of designing
serious games for refactoring on real-world
code artifacts. Specifically, they proposed a game design where students can compete either against a predefined
benchmark (technical debt) or against each other. In a follow-up work, Haendler \etal \cite{haendler2019refactutor,haendler2019interactive} developed an interactive
tutoring system for training software refactoring. Keuning \etal \cite{keuning2020student,keuning2021tutoring} to teach students to
refactor functionally correct code. More recently, Izu \etal \cite{izu2022resource} proposed a lab-based resource
to help novices identify and refactor antipatterns when writing
conditional statements.}   Although there are recent studies that explored refactoring practice in education in a general context, to the best
of our knowledge,  no prior studies have exposed the students to the process of \textit{detecting} bad programming practices, and \textit{correcting} them through applying suitable refactorings.  

\section{Conclusion and Future Work}
\label{Section:Conclusion}

In this study, we performed an experiment to understand how students perform the application of refactoring. Specifically, we demonstrate running an assignment about using the tool JDeodorant to refactor antipatterns. Overall, the participants rated various aspects of the plugin highly, while also providing valuable ideas for future development. We envision our findings enabling educators to support students with refactoring tools tuned towards safer refactoring. Future work in this area includes investigating students' understanding of refactoring using various real-world applications in a semester-long course project.  This offers the opportunity
for students to choose a refactoring-related topic.

\bibliographystyle{ACM-Reference-Format}
\bibliography{sample-base}


\begin{thebibliography}{36}


\ifx \showCODEN    \undefined \def \showCODEN     #1{\unskip}     \fi
\ifx \showDOI      \undefined \def \showDOI       #1{#1}\fi
\ifx \showISBNx    \undefined \def \showISBNx     #1{\unskip}     \fi
\ifx \showISBNxiii \undefined \def \showISBNxiii  #1{\unskip}     \fi
\ifx \showISSN     \undefined \def \showISSN      #1{\unskip}     \fi
\ifx \showLCCN     \undefined \def \showLCCN      #1{\unskip}     \fi
\ifx \shownote     \undefined \def \shownote      #1{#1}          \fi
\ifx \showarticletitle \undefined \def \showarticletitle #1{#1}   \fi
\ifx \showURL      \undefined \def \showURL       {\relax}        \fi
\providecommand\bibfield[2]{#2}
\providecommand\bibinfo[2]{#2}
\providecommand\natexlab[1]{#1}
\providecommand\showeprint[2][]{arXiv:#2}

\bibitem[\protect\citeauthoryear{??}{JDe}{[n.\,d.]}]%
        {JDeodorant-replication}
 \bibinfo{year}{[n.\,d.]}\natexlab{}.
\newblock
  \bibinfo{howpublished}{\url{https://anonymousresearchersubmission.github.io/SIGCSE2023_JDeodorant/}}.
\newblock


\bibitem[\protect\citeauthoryear{??}{Log}{[n.\,d.]}]%
        {Log4J}
 \bibinfo{year}{[n.\,d.]}\natexlab{}.
\newblock
  \bibinfo{howpublished}{\url{https://github.com/apache/logging-log4j2}}.
\newblock


\bibitem[\protect\citeauthoryear{Abid, Abdul~Basit, and Arshad}{Abid
  et~al\mbox{.}}{2015}]%
        {abid2015reflections}
\bibfield{author}{\bibinfo{person}{Shamsa Abid}, \bibinfo{person}{Hamid
  Abdul~Basit}, {and} \bibinfo{person}{Naveed Arshad}.}
  \bibinfo{year}{2015}\natexlab{}.
\newblock \showarticletitle{Reflections on teaching refactoring: A tale of two
  projects}. In \bibinfo{booktitle}{\emph{Proceedings of the 2015 ACM
  Conference on Innovation and Technology in Computer Science Education}}.
  \bibinfo{pages}{225--230}.
\newblock


\bibitem[\protect\citeauthoryear{AlOmar, AlOmar, and Mkaouer}{AlOmar
  et~al\mbox{.}}{2023}]%
        {alomar2023use}
\bibfield{author}{\bibinfo{person}{Eman~Abdullah AlOmar},
  \bibinfo{person}{Salma~Abdullah AlOmar}, {and} \bibinfo{person}{Mohamed~Wiem
  Mkaouer}.} \bibinfo{year}{2023}\natexlab{}.
\newblock \showarticletitle{On the use of static analysis to engage students
  with software quality improvement: An experience with pmd}.
\newblock \bibinfo{journal}{\emph{arXiv preprint arXiv:2302.05554}}
  (\bibinfo{year}{2023}).
\newblock


\bibitem[\protect\citeauthoryear{AlOmar, Chouchen, Mkaouer, and Ouni}{AlOmar
  et~al\mbox{.}}{2022}]%
        {alomar2022code}
\bibfield{author}{\bibinfo{person}{Eman~Abdullah AlOmar},
  \bibinfo{person}{Moataz Chouchen}, \bibinfo{person}{Mohamed~Wiem Mkaouer},
  {and} \bibinfo{person}{Ali Ouni}.} \bibinfo{year}{2022}\natexlab{}.
\newblock \showarticletitle{Code Review Practices for Refactoring Changes: An
  Empirical Study on OpenStack}.
\newblock  (\bibinfo{year}{2022}), \bibinfo{pages}{1--13}.
\newblock


\bibitem[\protect\citeauthoryear{AlOmar, Mkaouer, Newman, and Ouni}{AlOmar
  et~al\mbox{.}}{2021}]%
        {alomar2021preserving}
\bibfield{author}{\bibinfo{person}{Eman~Abdullah AlOmar},
  \bibinfo{person}{Mohamed~Wiem Mkaouer}, \bibinfo{person}{Christian Newman},
  {and} \bibinfo{person}{Ali Ouni}.} \bibinfo{year}{2021}\natexlab{}.
\newblock \showarticletitle{On preserving the behavior in software refactoring:
  A systematic mapping study}.
\newblock \bibinfo{journal}{\emph{Information and Software Technology}}
  (\bibinfo{year}{2021}), \bibinfo{pages}{106675}.
\newblock


\bibitem[\protect\citeauthoryear{Anquetil, Etien, Andreo, and Ducasse}{Anquetil
  et~al\mbox{.}}{2019}]%
        {anquetil2019decomposing}
\bibfield{author}{\bibinfo{person}{Nicolas Anquetil}, \bibinfo{person}{Anne
  Etien}, \bibinfo{person}{Gaelle Andreo}, {and} \bibinfo{person}{St{\'e}phane
  Ducasse}.} \bibinfo{year}{2019}\natexlab{}.
\newblock \showarticletitle{Decomposing god classes at siemens}. In
  \bibinfo{booktitle}{\emph{2019 IEEE International Conference on Software
  Maintenance and Evolution (ICSME)}}. IEEE, \bibinfo{pages}{169--180}.
\newblock


\bibitem[\protect\citeauthoryear{Ash and Clayton}{Ash and Clayton}{2009}]%
        {ash2009generating}
\bibfield{author}{\bibinfo{person}{Sarah~L Ash} {and} \bibinfo{person}{Patti~H
  Clayton}.} \bibinfo{year}{2009}\natexlab{}.
\newblock \showarticletitle{Generating, deepening, and documenting learning:
  The power of critical reflection in applied learning}.
\newblock  (\bibinfo{year}{2009}).
\newblock


\bibitem[\protect\citeauthoryear{Bavota, De~Lucia, Di~Penta, Oliveto, and
  Palomba}{Bavota et~al\mbox{.}}{2015}]%
        {bavota2015experimental}
\bibfield{author}{\bibinfo{person}{Gabriele Bavota}, \bibinfo{person}{Andrea
  De~Lucia}, \bibinfo{person}{Massimiliano Di~Penta}, \bibinfo{person}{Rocco
  Oliveto}, {and} \bibinfo{person}{Fabio Palomba}.}
  \bibinfo{year}{2015}\natexlab{}.
\newblock \showarticletitle{An experimental investigation on the innate
  relationship between quality and refactoring}.
\newblock \bibinfo{journal}{\emph{Journal of Systems and Software}}
  \bibinfo{volume}{107} (\bibinfo{year}{2015}), \bibinfo{pages}{1--14}.
\newblock


\bibitem[\protect\citeauthoryear{Berland, Martin, Ko, Peacock, Rudolph, and
  Golubski}{Berland et~al\mbox{.}}{2013}]%
        {berland2013student}
\bibfield{author}{\bibinfo{person}{Leema~K Berland}, \bibinfo{person}{Taylor~H
  Martin}, \bibinfo{person}{Pat Ko}, \bibinfo{person}{Stephanie~Baker Peacock},
  \bibinfo{person}{Jennifer~J Rudolph}, {and} \bibinfo{person}{Chris
  Golubski}.} \bibinfo{year}{2013}\natexlab{}.
\newblock \showarticletitle{Student learning in challenge-based engineering
  curricula}.
\newblock \bibinfo{journal}{\emph{Journal of Pre-College Engineering Education
  Research (J-PEER)}} \bibinfo{volume}{3}, \bibinfo{number}{1}
  (\bibinfo{year}{2013}), \bibinfo{pages}{5}.
\newblock


\bibitem[\protect\citeauthoryear{Bessghaier, Ouni, and Mkaouer}{Bessghaier
  et~al\mbox{.}}{2021}]%
        {bessghaier2021longitudinal}
\bibfield{author}{\bibinfo{person}{Narjes Bessghaier}, \bibinfo{person}{Ali
  Ouni}, {and} \bibinfo{person}{Mohamed~Wiem Mkaouer}.}
  \bibinfo{year}{2021}\natexlab{}.
\newblock \showarticletitle{A longitudinal exploratory study on code smells in
  server side web applications}.
\newblock \bibinfo{journal}{\emph{Software Quality Journal}}
  \bibinfo{volume}{29}, \bibinfo{number}{4} (\bibinfo{year}{2021}),
  \bibinfo{pages}{901--941}.
\newblock


\bibitem[\protect\citeauthoryear{Brockbank and McGill}{Brockbank and
  McGill}{2007}]%
        {brockbank2007facilitating}
\bibfield{author}{\bibinfo{person}{Anne Brockbank} {and} \bibinfo{person}{Ian
  McGill}.} \bibinfo{year}{2007}\natexlab{}.
\newblock \bibinfo{booktitle}{\emph{Facilitating reflective learning in higher
  education}}.
\newblock \bibinfo{publisher}{McGraw-Hill Education (UK)}.
\newblock


\bibitem[\protect\citeauthoryear{Chidamber and Kemerer}{Chidamber and
  Kemerer}{1994}]%
        {chidamber1994metrics}
\bibfield{author}{\bibinfo{person}{Shyam~R Chidamber} {and}
  \bibinfo{person}{Chris~F Kemerer}.} \bibinfo{year}{1994}\natexlab{}.
\newblock \showarticletitle{A metrics suite for object oriented design}.
\newblock \bibinfo{journal}{\emph{IEEE Transactions on software engineering}}
  \bibinfo{volume}{20}, \bibinfo{number}{6} (\bibinfo{year}{1994}),
  \bibinfo{pages}{476--493}.
\newblock


\bibitem[\protect\citeauthoryear{Colbeck, Campbell, and Bjorklund}{Colbeck
  et~al\mbox{.}}{2000}]%
        {colbeck2000grouping}
\bibfield{author}{\bibinfo{person}{Carol~L Colbeck}, \bibinfo{person}{Susan~E
  Campbell}, {and} \bibinfo{person}{Stefani~A Bjorklund}.}
  \bibinfo{year}{2000}\natexlab{}.
\newblock \showarticletitle{Grouping in the dark: What college students learn
  from group projects}.
\newblock \bibinfo{journal}{\emph{The Journal of Higher Education}}
  \bibinfo{volume}{71}, \bibinfo{number}{1} (\bibinfo{year}{2000}),
  \bibinfo{pages}{60--83}.
\newblock


\bibitem[\protect\citeauthoryear{Cruzes and Dyba}{Cruzes and Dyba}{2011}]%
        {cruzes2011recommended}
\bibfield{author}{\bibinfo{person}{Daniela~S Cruzes} {and}
  \bibinfo{person}{Tore Dyba}.} \bibinfo{year}{2011}\natexlab{}.
\newblock \showarticletitle{Recommended steps for thematic synthesis in
  software engineering}. In \bibinfo{booktitle}{\emph{2011 international
  symposium on empirical software engineering and measurement}}. IEEE,
  \bibinfo{pages}{275--284}.
\newblock


\bibitem[\protect\citeauthoryear{Cunningham}{Cunningham}{1992}]%
        {cunningham1992wycash}
\bibfield{author}{\bibinfo{person}{Ward Cunningham}.}
  \bibinfo{year}{1992}\natexlab{}.
\newblock \showarticletitle{The WyCash portfolio management system}.
\newblock \bibinfo{journal}{\emph{ACM SIGPLAN OOPS Messenger}}
  \bibinfo{volume}{4}, \bibinfo{number}{2} (\bibinfo{year}{1992}),
  \bibinfo{pages}{29--30}.
\newblock


\bibitem[\protect\citeauthoryear{Elezi, Sali, Demeyer, Murgia, and
  P{\'e}rez}{Elezi et~al\mbox{.}}{2016}]%
        {elezi2016game}
\bibfield{author}{\bibinfo{person}{Leonard Elezi}, \bibinfo{person}{Sara Sali},
  \bibinfo{person}{Serge Demeyer}, \bibinfo{person}{Alessandro Murgia}, {and}
  \bibinfo{person}{Javier P{\'e}rez}.} \bibinfo{year}{2016}\natexlab{}.
\newblock \showarticletitle{A game of refactoring: Studying the impact of
  gamification in software refactoring}. In
  \bibinfo{booktitle}{\emph{Proceedings of the Scientific Workshop Proceedings
  of XP2016}}. \bibinfo{pages}{1--6}.
\newblock


\bibitem[\protect\citeauthoryear{Fowler, Beck, Brant, Opdyke, and
  Roberts}{Fowler et~al\mbox{.}}{1999}]%
        {Fowler:1999:RID:311424}
\bibfield{author}{\bibinfo{person}{Martin Fowler}, \bibinfo{person}{Kent Beck},
  \bibinfo{person}{John Brant}, \bibinfo{person}{William Opdyke}, {and}
  \bibinfo{person}{don Roberts}.} \bibinfo{year}{1999}\natexlab{}.
\newblock \bibinfo{booktitle}{\emph{Refactoring: Improving the Design of
  Existing Code}}.
\newblock \bibinfo{publisher}{Addison-Wesley Longman Publishing Co., Inc.},
  \bibinfo{address}{Boston, MA, USA}.
\newblock
\showISBNx{0-201-48567-2}
\urldef\tempurl%
\url{http://dl.acm.org/citation.cfm?id=311424}
\showURL{%
\tempurl}


\bibitem[\protect\citeauthoryear{Golubev, Kurbatova, AlOmar, Bryksin, and
  Mkaouer}{Golubev et~al\mbox{.}}{2021}]%
        {golubev2021one}
\bibfield{author}{\bibinfo{person}{Yaroslav Golubev}, \bibinfo{person}{Zarina
  Kurbatova}, \bibinfo{person}{Eman~Abdullah AlOmar}, \bibinfo{person}{Timofey
  Bryksin}, {and} \bibinfo{person}{Mohamed~Wiem Mkaouer}.}
  \bibinfo{year}{2021}\natexlab{}.
\newblock \showarticletitle{One thousand and one stories: a large-scale survey
  of software refactoring}. In \bibinfo{booktitle}{\emph{Proceedings of the
  29th ACM Joint Meeting on European Software Engineering Conference and
  Symposium on the Foundations of Software Engineering}}.
  \bibinfo{pages}{1303--1313}.
\newblock


\bibitem[\protect\citeauthoryear{Grissom and Kim}{Grissom and Kim}{2005}]%
        {trove.nla.gov.au/work/16432558}
\bibfield{author}{\bibinfo{person}{Robert~J Grissom} {and}
  \bibinfo{person}{John~J Kim}.} \bibinfo{year}{2005}\natexlab{}.
\newblock \bibinfo{booktitle}{\emph{Effect sizes for research : a broad
  practical approach}}.
\newblock \bibinfo{publisher}{Mahwah, N.J. ; London : Lawrence Erlbaum
  Associates}.
\newblock
\showISBNx{0805850147 (alk. paper)}
\newblock
\shownote{Formerly CIP}.


\bibitem[\protect\citeauthoryear{Haendler and Neumann}{Haendler and
  Neumann}{2019}]%
        {haendler2019serious}
\bibfield{author}{\bibinfo{person}{Thorsten Haendler} {and}
  \bibinfo{person}{Gustaf Neumann}.} \bibinfo{year}{2019}\natexlab{}.
\newblock \showarticletitle{Serious refactoring games}. In
  \bibinfo{booktitle}{\emph{Proceedings of the 52nd Hawaii International
  Conference on System Sciences}}.
\newblock


\bibitem[\protect\citeauthoryear{Haendler, Neumann, and Smirnov}{Haendler
  et~al\mbox{.}}{2019a}]%
        {haendler2019interactive}
\bibfield{author}{\bibinfo{person}{Thorsten Haendler}, \bibinfo{person}{Gustaf
  Neumann}, {and} \bibinfo{person}{Fiodor Smirnov}.}
  \bibinfo{year}{2019}\natexlab{a}.
\newblock \showarticletitle{An interactive tutoring system for training
  software refactoring}.
\newblock \bibinfo{journal}{\emph{Instructor}}  \bibinfo{volume}{1}
  (\bibinfo{year}{2019}), \bibinfo{pages}{4}.
\newblock


\bibitem[\protect\citeauthoryear{Haendler, Neumann, and Smirnov}{Haendler
  et~al\mbox{.}}{2019b}]%
        {haendler2019refactutor}
\bibfield{author}{\bibinfo{person}{Thorsten Haendler}, \bibinfo{person}{Gustaf
  Neumann}, {and} \bibinfo{person}{Fiodor Smirnov}.}
  \bibinfo{year}{2019}\natexlab{b}.
\newblock \showarticletitle{RefacTutor: an interactive tutoring system for
  software refactoring}. In \bibinfo{booktitle}{\emph{International Conference
  on Computer Supported Education}}. Springer, \bibinfo{pages}{236--261}.
\newblock


\bibitem[\protect\citeauthoryear{Izu, Denny, and Roy}{Izu
  et~al\mbox{.}}{2022}]%
        {izu2022resource}
\bibfield{author}{\bibinfo{person}{Cruz Izu}, \bibinfo{person}{Paul Denny},
  {and} \bibinfo{person}{Sayoni Roy}.} \bibinfo{year}{2022}\natexlab{}.
\newblock \showarticletitle{A Resource to Support Novices Refactoring
  Conditional Statements}. In \bibinfo{booktitle}{\emph{Proceedings of the 27th
  ACM Conference on on Innovation and Technology in Computer Science Education
  Vol. 1}}. \bibinfo{pages}{344--350}.
\newblock


\bibitem[\protect\citeauthoryear{Keuning, Heeren, and Jeuring}{Keuning
  et~al\mbox{.}}{2020}]%
        {keuning2020student}
\bibfield{author}{\bibinfo{person}{Hieke Keuning}, \bibinfo{person}{Bastiaan
  Heeren}, {and} \bibinfo{person}{Johan Jeuring}.}
  \bibinfo{year}{2020}\natexlab{}.
\newblock \showarticletitle{Student refactoring behaviour in a programming
  tutor}. In \bibinfo{booktitle}{\emph{Koli Calling'20: Proceedings of the 20th
  Koli Calling International Conference on Computing Education Research}}.
  \bibinfo{pages}{1--10}.
\newblock


\bibitem[\protect\citeauthoryear{Keuning, Heeren, and Jeuring}{Keuning
  et~al\mbox{.}}{2021}]%
        {keuning2021tutoring}
\bibfield{author}{\bibinfo{person}{Hieke Keuning}, \bibinfo{person}{Bastiaan
  Heeren}, {and} \bibinfo{person}{Johan Jeuring}.}
  \bibinfo{year}{2021}\natexlab{}.
\newblock \showarticletitle{A tutoring system to learn code refactoring}. In
  \bibinfo{booktitle}{\emph{Proceedings of the 52nd ACM Technical Symposium on
  Computer Science Education}}. \bibinfo{pages}{562--568}.
\newblock


\bibitem[\protect\citeauthoryear{Khomh, Di~Penta, and Gueheneuc}{Khomh
  et~al\mbox{.}}{2009}]%
        {khomh2009exploratory}
\bibfield{author}{\bibinfo{person}{Foutse Khomh}, \bibinfo{person}{Massimiliano
  Di~Penta}, {and} \bibinfo{person}{Yann-Gael Gueheneuc}.}
  \bibinfo{year}{2009}\natexlab{}.
\newblock \showarticletitle{An exploratory study of the impact of code smells
  on software change-proneness}. In \bibinfo{booktitle}{\emph{2009 16th Working
  Conference on Reverse Engineering}}. IEEE, \bibinfo{pages}{75--84}.
\newblock


\bibitem[\protect\citeauthoryear{Khomh, Penta, Gu{\'e}h{\'e}neuc, and
  Antoniol}{Khomh et~al\mbox{.}}{2012}]%
        {khomh2012exploratory}
\bibfield{author}{\bibinfo{person}{Foutse Khomh},
  \bibinfo{person}{Massimiliano~Di Penta}, \bibinfo{person}{Yann-Ga{\"e}l
  Gu{\'e}h{\'e}neuc}, {and} \bibinfo{person}{Giuliano Antoniol}.}
  \bibinfo{year}{2012}\natexlab{}.
\newblock \showarticletitle{An exploratory study of the impact of antipatterns
  on class change-and fault-proneness}.
\newblock \bibinfo{journal}{\emph{Empirical Software Engineering}}
  \bibinfo{volume}{17}, \bibinfo{number}{3} (\bibinfo{year}{2012}),
  \bibinfo{pages}{243--275}.
\newblock


\bibitem[\protect\citeauthoryear{L{\'o}pez, Alonso, Marticorena, and
  Maudes}{L{\'o}pez et~al\mbox{.}}{2014}]%
        {lopez2014design}
\bibfield{author}{\bibinfo{person}{Carlos L{\'o}pez},
  \bibinfo{person}{Jes{\'u}s~M Alonso}, \bibinfo{person}{Ra{\'u}l Marticorena},
  {and} \bibinfo{person}{Jes{\'u}s~M Maudes}.} \bibinfo{year}{2014}\natexlab{}.
\newblock \showarticletitle{Design of e-activities for the learning of code
  refactoring tasks}. In \bibinfo{booktitle}{\emph{2014 International Symposium
  on Computers in Education (SIIE)}}. IEEE, \bibinfo{pages}{35--40}.
\newblock


\bibitem[\protect\citeauthoryear{Palomba, Bavota, Di~Penta, Fasano, Oliveto,
  and De~Lucia}{Palomba et~al\mbox{.}}{2018}]%
        {palomba2018diffuseness}
\bibfield{author}{\bibinfo{person}{Fabio Palomba}, \bibinfo{person}{Gabriele
  Bavota}, \bibinfo{person}{Massimiliano Di~Penta}, \bibinfo{person}{Fausto
  Fasano}, \bibinfo{person}{Rocco Oliveto}, {and} \bibinfo{person}{Andrea
  De~Lucia}.} \bibinfo{year}{2018}\natexlab{}.
\newblock \showarticletitle{On the diffuseness and the impact on
  maintainability of code smells: a large scale empirical investigation}.
\newblock \bibinfo{journal}{\emph{Empirical Software Engineering}}
  \bibinfo{volume}{23}, \bibinfo{number}{3} (\bibinfo{year}{2018}),
  \bibinfo{pages}{1188--1221}.
\newblock


\bibitem[\protect\citeauthoryear{Raab}{Raab}{2012}]%
        {raab2012codesmellexplorer}
\bibfield{author}{\bibinfo{person}{Felix Raab}.}
  \bibinfo{year}{2012}\natexlab{}.
\newblock \showarticletitle{CodeSmellExplorer: Tangible exploration of code
  smells and refactorings}. In \bibinfo{booktitle}{\emph{2012 IEEE Symposium on
  Visual Languages and Human-Centric Computing (VL/HCC)}}. IEEE,
  \bibinfo{pages}{261--262}.
\newblock


\bibitem[\protect\citeauthoryear{Silva, Tsantalis, and Valente}{Silva
  et~al\mbox{.}}{2016}]%
        {Silva:2016:WWR:2950290.2950305}
\bibfield{author}{\bibinfo{person}{Danilo Silva}, \bibinfo{person}{Nikolaos
  Tsantalis}, {and} \bibinfo{person}{Marco~Tulio Valente}.}
  \bibinfo{year}{2016}\natexlab{}.
\newblock \showarticletitle{Why We Refactor? Confessions of GitHub
  Contributors}. In \bibinfo{booktitle}{\emph{Proceedings of the 2016 24th ACM
  SIGSOFT International Symposium on Foundations of Software Engineering}}
  (Seattle, WA, USA) \emph{(\bibinfo{series}{FSE 2016})}.
  \bibinfo{publisher}{ACM}, \bibinfo{address}{New York, NY, USA},
  \bibinfo{pages}{858--870}.
\newblock
\showISBNx{978-1-4503-4218-6}
\urldef\tempurl%
\url{https://doi.org/10.1145/2950290.2950305}
\showDOI{\tempurl}


\bibitem[\protect\citeauthoryear{Smith, Stoecklin, and Serino}{Smith
  et~al\mbox{.}}{2006}]%
        {smith2006innovative}
\bibfield{author}{\bibinfo{person}{Suzanne Smith}, \bibinfo{person}{Sara
  Stoecklin}, {and} \bibinfo{person}{Catharina Serino}.}
  \bibinfo{year}{2006}\natexlab{}.
\newblock \showarticletitle{An innovative approach to teaching refactoring}. In
  \bibinfo{booktitle}{\emph{Proceedings of the 37th SIGCSE technical symposium
  on Computer science education}}. \bibinfo{pages}{349--353}.
\newblock


\bibitem[\protect\citeauthoryear{Stoecklin, Smith, and Serino}{Stoecklin
  et~al\mbox{.}}{2007}]%
        {stoecklin2007teaching}
\bibfield{author}{\bibinfo{person}{Sara Stoecklin}, \bibinfo{person}{Suzanne
  Smith}, {and} \bibinfo{person}{Catharina Serino}.}
  \bibinfo{year}{2007}\natexlab{}.
\newblock \showarticletitle{Teaching students to build well formed
  object-oriented methods through refactoring}.
\newblock \bibinfo{journal}{\emph{ACM SIGCSE Bulletin}} \bibinfo{volume}{39},
  \bibinfo{number}{1} (\bibinfo{year}{2007}), \bibinfo{pages}{145--149}.
\newblock


\bibitem[\protect\citeauthoryear{Tsantalis, Chaikalis, and
  Chatzigeorgiou}{Tsantalis et~al\mbox{.}}{2018}]%
        {tsantalis2018ten}
\bibfield{author}{\bibinfo{person}{Nikolaos Tsantalis},
  \bibinfo{person}{Theodoros Chaikalis}, {and} \bibinfo{person}{Alexander
  Chatzigeorgiou}.} \bibinfo{year}{2018}\natexlab{}.
\newblock \showarticletitle{Ten years of JDeodorant: Lessons learned from the
  hunt for smells}. In \bibinfo{booktitle}{\emph{2018 IEEE 25th international
  conference on software analysis, evolution and reengineering (SANER)}}. IEEE,
  \bibinfo{pages}{4--14}.
\newblock


\bibitem[\protect\citeauthoryear{Wilcoxon}{Wilcoxon}{1945}]%
        {wilcoxon1945individual}
\bibfield{author}{\bibinfo{person}{Frank Wilcoxon}.}
  \bibinfo{year}{1945}\natexlab{}.
\newblock \showarticletitle{Individual comparisons by ranking methods}.
\newblock \bibinfo{journal}{\emph{Biometrics bulletin}} \bibinfo{volume}{1},
  \bibinfo{number}{6} (\bibinfo{year}{1945}), \bibinfo{pages}{80--83}.
\newblock


\end{thebibliography}

\end{document}